\documentclass{article}
\usepackage[final]{neurips_2021}
\usepackage[utf8]{inputenc}

\usepackage[T1]{fontenc}    % use 8-bit T1 fonts
\usepackage{hyperref}       % hyperlinks
\usepackage{url}            % simple URL typesetting
\usepackage{booktabs}       % professional-quality tables
\usepackage{amsfonts}       % blackboard math symbols
\usepackage{nicefrac}       % compact symbols for 1/2, etc.
\usepackage{microtype}      % microtypography
\usepackage{xcolor}         % colors
\usepackage{natbib}
\usepackage{graphicx}
\usepackage[font=scriptsize]{subcaption}
\makeatletter
\newcommand{\printfnsymbol}[1]{%
  \textsuperscript{\@fnsymbol{#1}}%
}
\makeatother

\title{Automatic Detection of Alzheimer’s Disease with Multi-Modal Fusion of Clinical MRI Scans}
\author{
    Long Chen\\ Center for Data Science \\ New York University \\ \texttt{lc3424@nyu.edu} \\
    \And Liben Chen\\ Center for Data Science \\ New York University \\ \texttt{lc4438@nyu.edu} \\
    \And Binfeng Xu\\ Center for Data Science \\ New York University \\ \texttt{bx2010@nyu.edu}\\
    \AND Wenxin Zhang\\ Center for Data Science \\ New York University \\ \texttt{wz2164@nyu.edu} \\
    \And Narges Razavian \\ NYU Grossman School of Medicine \\and NYU Center for Data Science \\ \texttt{Narges.Razavian@nyulangone.org}
}
\date{December 2021}

\begin{document}

\maketitle

\section{Introduction}

% Motivate and abstractly describe the problem you are solving and how you are addressing it. What is the problem? Why is it important? What is your basic approach? A short discussion of how it fits into related work in the area is also desirable. Summarize the basic results and conclusions that you will present.

The aging population of the U.S. drives the prevalence of Alzheimer's disease. \citet{brookmeyer2018forecasting} forecasts approximately 15 million Americans will have either clinical AD or mild cognitive impairment by 2060. In response to this urgent call, methods for early detection of Alzheimer's disease have been developed for prevention and pre-treatment. Notably, literature on the application of deep learning in the automatic detection of the disease has been proliferating \citep{noor2020application}. This study builds upon previous literature and maintains a focus on leveraging multi-modal information to enhance automatic detection. We aim to predict the stage of the disease - Cognitively Normal (CN), Mildly Cognitive Impairment (MCI), and Alzheimer's Disease (AD), based on two different types of brain MRI scans. We design an AlexNet-based deep learning model that learns the synergy of complementary information from both T1 and FLAIR MRI scans. The contributions of our study are
\begin{itemize}
    \item This is the first study on automatic detection of Alzheimer's disease with \textbf{a clinical data set}. Previous literature builds on an experimental dataset from ADNI, while we conduct experiments on a real-life clinical dataset from NYU Barlow. 
    \item This study empirically evidences \textbf{feasibility} of transfer learning on dementia detection. The model trained on the experimental data set retains useful knowledge that can be readily transferred to learning tasks on the clinical data set.  
    \item We deliver three \textbf{competitive} baseline models that can be benchmarked against models in transfer learning and multi-modal learning task on detection of dementia. The best model achieves \textbf{a micro-AUC score} $\mathbf{0.88}$.
\end{itemize}

\section{Related Work}

\subsection{Deep multi-modal Learning}
Deep multi-modal learning concerns how information from different sources of input can be fused together to improve the quality of the learned representation. Previous literature in deep multi-modal learning mostly focuses on the task of semantic segmentation \citep{cheng2017locality, hazirbas2016fusenet,ngiam2011multimodal} in computer vision study, using a data set featuring two types of input - the RGB color image and the greyscale depth image. Many effective methods are developed and most of them fall into two categories - aggregation-based and alignment-based multi-modal fusion.

The aggregation-based multi-modal fusion applies averaging \citep{hazirbas2016fusenet}, concatenation \citep{ngiam2011multimodal}, and self-attention \citep{zeng2019deep} operation over pair of vector representations from two modalities. Previous literature has shown the simple averaging and concatenation between vector representation from two modalities can already improve the model performance by a substantial margin \citep{hazirbas2016fusenet, ngiam2011multimodal}. With more sophisticated operations such self-attention, \cite{zeng2019deep} designs a model that uses information from other modalities to adaptively select useful features in the focal modality. This too evidences that a principled way of multi-modal fusion can enhance the final representation learned by the model. 

The alignment-based multi-modal fusion \citep{song2020modality, cheng2017locality, wang2016learning} tries to align the common information that is presented across modalities and also to extract distinctive information in each of the modality. \citet{cheng2017locality} designs a model where the representation from each modality is linearly projected to two subsequent vectors - one retains information that is unique to its own modality and the other retains information that is common to other modalities. They designs a loss function in a way such that the Multi-kernel Maximum Mean Distance (MK-MMD) between vectors that retain common information across modalities should be minimized and the MK-MMD between vectors that retain unique information to its own modality should be maximized. By aligning these vectors in this way, the model is enabled to select information from different modalities that are complementary to each other, improving the quality of the final representation and the model performance.

\subsection{Multi-modal learning on automatic detection of alzheimer's diseases}

Previous studies \citep{liu2018multi, vu2017multimodal, liu2014multimodal,  liu2014inter, zhang2011multimodal} mostly focus on using MRI scans and PET scans as two modalities in the detection of Alzheimer's diseases. \citet{zhang2011multimodal, liu2014inter} builds kernel matrices for features selected from MRI, PET, and CSF data, and combines kernels for Support Vector Machine classification. \citet{liu2018multi} applies CNN in extracting features from MRI and PET brain scans and fuses the representations from Convolutional Neural Network (CNN) \citep{lecun1995convolutional} for final predictions. Later, with advancement in self-supervised learning, \citet{vu2017multimodal} applies Sparse Auto-encoder (SAE) \citep{ng2011sparse} in pretraining embedding from MRI and PET brain scans and pipes the pre-trained embedding into CNN for classification.

This study differs from previous ones in that we focus on multi-modal MRI images, namely T1 and FLAIR scans. We don't use PET scans as a modality in our study. Also, we don't use any sophisticated multi-modal fusion techniques in our study. We only use a basic CNN architecture and apply ensemble learning over predictions from models trained separately on each modality. Our evaluation result shows that, with proper parameter tuning, this simple design of architecture can still achieve competitive and even superior performance than more complex models in many previous studies. 

% Brain MRI scans consist of different modalities - T1, T2, and FLAIR. Each of modality is a type of scan that highlights different types of tissues in the brain. Intuitively, by learning on different types of brain MRI images 

\section{Problem Definition and Algorithm}

\subsection{Problem background}

In this study, we address the problem of classifying stage of the Alzheimer's disease based on T1 and FLAIR scans. The motivation for using different types of MRI scans is that these scans provide different types of useful information that can aid in the diagnosis of Alzheimer's disease. For example, a T1 scan highlights the fat and white matter of the brain, while FLAIR highlights the inflammation and cerebrospinal fluid (CSF).  The previous study \citep{mcgeer1999brain} has shown that brain inflammation is a pathological cause of Alzheimer's disease. Another study \citep{mattsson2012age} has found CSF an important biomarker for the disease. Information on both CSF and inflammation can be readily extracted from the FLAIR scan and it is intuitive to include this modality in our predictive modeling for better detection performance. 

\subsection{Problem formulation}

\textbf{Stages of Disease} Following previous literature \citep{liu2021development, liu2020design}, we define three stages of Alzheimer's diseases - Cognitive Normal (CN), Mildly Cognitive Impairment (MCI), and Alzheimer's disease (AD). We denote the stage of the disease as a random variable $Y$. 

\textbf{Multimodal Inputs} We denote the feature matrix from the T1 scan as $X_1$ and the feature matrix from the FLAIR scan as $X_2$. 

Formally, the problem of automatic detection of Alzheimer's disease is to model the probability of the patient's stage of Alzheimer's disease 
\begin{equation}
    \mathbb{P}(Y \vert X_1, X_2)
\end{equation}

\subsection{Algorithm}

We adopt AlexNet-based CNN architecture in extracting features from the T1 and FLAIR images. The specification of the architecture is listed in the Table \ref{architect}. The input to architecture is a 3D T1 or FLAIR MRI image with a size of the dimension $96 \times 96 \times 96$. The architecture consists of 4 basic blocks, each of which is stacked with a 3D convolutional layer, an instanace normalisation layer, a ReLU layer, and a max-pooling layer. The original input image is transformed by four basic blocks and is output by the final block as a $5 \times 5 \times 5$ feature map. The feature map is then flattened as an input to the first fully connected layer that transforms it into a vector of the dimension $3 \times 1$. The softmax layer takes in this vector and outputs the probability for three classes at the end. We optimize this model against the cross-entropy loss. 

\begin{table}
  \caption{The AlexNet-based CNN Architecture}
  \label{architect}
  \centering
  \begin{tabular}{llll}
    \toprule
    % \multicolumn{2}{c}{Part}                   \\
    \cmidrule(r){1-2}
    Block     & Layer     & Type & Output size\\
    \midrule
             & Inputs     &     & $96 \times 96 \times 96$\\
    \midrule
    1 & Conv3D  & k1-c4$\cdot$f-p0-s1-d1 &$96 \times 96 \times 96$\\
      & InstanceNorm3D & &      \\
      & ReLU       &  &  \\
      & MaxPool3D  & k3-s2 &$47 \times 47 \times 47$\\
    2 & Conv3D  & k3-c32$\cdot$f-p0-s1-d2 &$43 \times 43 \times 43$\\
      & InstanceNorm3D & &      \\
      & ReLU       &  &  \\
      & MaxPool3D  & k3-s2 &$21 \times 21 \times 21$\\
    3 & Conv3D  & k5-c64$\cdot$f-p2-s1-d2 &$17 \times 17 \times 17$\\
      & InstanceNorm3D & &      \\
      & ReLU       &  &  \\
      & MaxPool3D  & k3-s2 &$8 \times 8 \times 8$\\
    4 & Conv3D  & k3-c64$\cdot$f-p1-s1-d2 &$6 \times 6 \times 6$\\
      & InstanceNorm3D & &      \\
      & ReLU       &  &  \\
      & MaxPool3D  & k5-s2 &$5 \times 5 \times 5$\\
    \midrule
    FC1 & & 1024 & \\ 
    FC2 & & 3 & \\ 
    Softmax & & 3 & \\ 
    \bottomrule
  \end{tabular}
    \vspace{.1cm}
  \newline
    {\raggedright \small Notes: k = kernel size, c = number of channels as a multiple of the widening factor f, p = padding size, s=stride, and d = dilation. We adapt this Table from \citet{liu2020design}.}
\end{table}

\section{Datasets and Preprocessing}
\subsection{Datasets}
For this study, we use MRI scans collected from NYU Pearl I. Barlow Center for Memory Evaluation \& Treatment (NYU Barlow), which are raw, clinical MRI images. The dataset contains both T1-weighted (T1) and FLAIR structural MRI scans. We generate a multi-modal dataset by identifying sessions with a good-quality. T1 and a FLAIR MRI image. We then utilize electronic health records (EHR) from NYU Barlow to identify labels for each scans. The labels are AD (mildly demented patients diagnosed with AD), MCI (mildly cognitively-impaired patients in the prodromal phase of AD) and CN (elderly control participants). We perform patient-level dataset split for training (70\%), validation (15\%) and testing (15\%) sets. Visualizations of preprocessing and labelling pipelines are shown in Figure.~\ref{fig:preprocessing_pipeline}.

\subsection{Image preprocessing}
Most previous studies use packages including FSL~\citep{FSL}, Statistical Parametric Mapping (SPM)~\citep{SPM}, and FreeSurfer~\citep{fischl2012freesurfer}. FSL provides brain extraction and tissue segmentation functionality, while SPM realigns, spatially normalizes, and smooths the scans. FreeSurfer provides a preprocessing stream that includes skull stripping, segmentation, and nonlinear registration. For this study, we utilized two pipelines to preprocess T1 and FLAIR image respectively.

\paragraph{T1-Weighted MRI Preprocessing} For T1 scans, we used the Clinica~\citep{Clinica} software platform developed by ARAMIS Lab, which supports FSL, SPM and FreeSurfer. First, we perform image centering and re-orientation with FSL in order to ensure rough spatial uniformity across all images. We then use Clinica to register the scans to a Dartel template computed exclusively from the training data~\citep{dartel}, and normalize them to the Montreal Neurological Institute (MNI) coordinate space~\citep{evans19933d}. The input to the Clinica software is the BIDS-formatted NYU Barlow T1 dataset. The output dimensions are $121 \times 145 \times 121$ voxels along sagittal, coronal and axial dimensions, respectively.

\paragraph{FLAIR MRI Preprocessing} For FLAIR scans, we also perform image centering and re-orientation with FSL. Next, FLIRT (FMRIB's Linear Image Registration Tool)~\citep{jenkinson2001global, jenkinson2002improved} is used to register the images. We utilize a preprocessed T1 image from the same scan session to serve as a reference, template image for the FLIRT pipeline. The input to the FLIRT algorithm is also the FLAIR scans in BIDS format. The output dimensions are the same with T1 preprocessed scans (i.e. $121 \times 145 \times 121$ voxels along sagittal, coronal and axial dimensions, respectively).

\subsection{Labelling pipeline}
We utilize NYU Barlow EHR as raw patient health records. A series of cleaning and processing steps is performed to achieve a labeled dataset. First, we retain only diagnoses where the patient is over 55 at the time of scan. Then, we derive label (AD, MCI or CN) for each visit according to the doctor's diagnosis from various evaluation metrics. Next, re remove entries with no diagnosed label and entries that are 180 days prior to or later than the scan date. In addition, we check for temporal consistency for each patient, ensuring that patients do not have milder diagnoses after a more severe one. For example, patients diagnosed as AD cannot be diagnosed with MCI or CN at a later date, and those who diagnosed as MCI cannot go back to CN. All less severe diagnoses are corrected to the previous, more severe label. We then take the mode diagnosis as the label for each scan, with tie-breaking rule preferring the worst diagnosis among the three label candidates.

\bigskip
Due to errors in image preprocessing processes and missing or bad annotations during the labelling pipeline, the final number of avaiable T1-Flair-Label entries is 3095. \textcolor{red}{Table X} shows the statistics of the patients in the training, validation and testing sets.

\section{Experiment and Results}
\subsection{Evaluation Metrics}
To evaluate performance of the classifier that we build, we adopt common criteria for multi-class classification task, namely, AUC (Area under Curve), Precision, and Recall.

\subsection{Description of computational experiments}
Our experiment can be divided into two main categories. First, we discover the feasibility of transfer learning. The two transfer learning tasks are T1-T1 from different datasets and T1-FLAIR within the same dataset. Second, we discover the fusion of multi-modal MRIs for better performance.

\subsubsection{T1-T1 transfer learning}
For the T1-T1 transfer learning task, we aim to check the performance of using pre-trained model~\citep{liu2020design} trained with experimental T1-weighted scans from ADNI~\citep{ADNI} dataset onto our clinical T1-weighted dataset. Even with same modality, scan quality and intensity can differ due to discrepancies in equipment and on-device image processing algorithms. First, we directly apply the pre-trained model to our testing set as our baseline model. Next, we fine-tune the fully connected layers with training set. Finally, we re-train the model with training set. We kept the model architecture identical in these experiments.

We perform data augmentation via Gaussian blurring with $\sigma$ uniformly chosen from 0 to 1.5, and random cropping of size $96 \times 96 \times 96$. For fine-tuning, we set batch size to 4, image encoding dimension to 1024 and learning rate to 0.01. We use stochastic gradient descent with momentum equal to 0.9. We train the model for 50 epochs. For re-training, we use the same hyperparameters and training settings, but with longer training epoch of 200.

\begin{figure}[t]
    \centering
    \includegraphics[width=.95\textwidth]{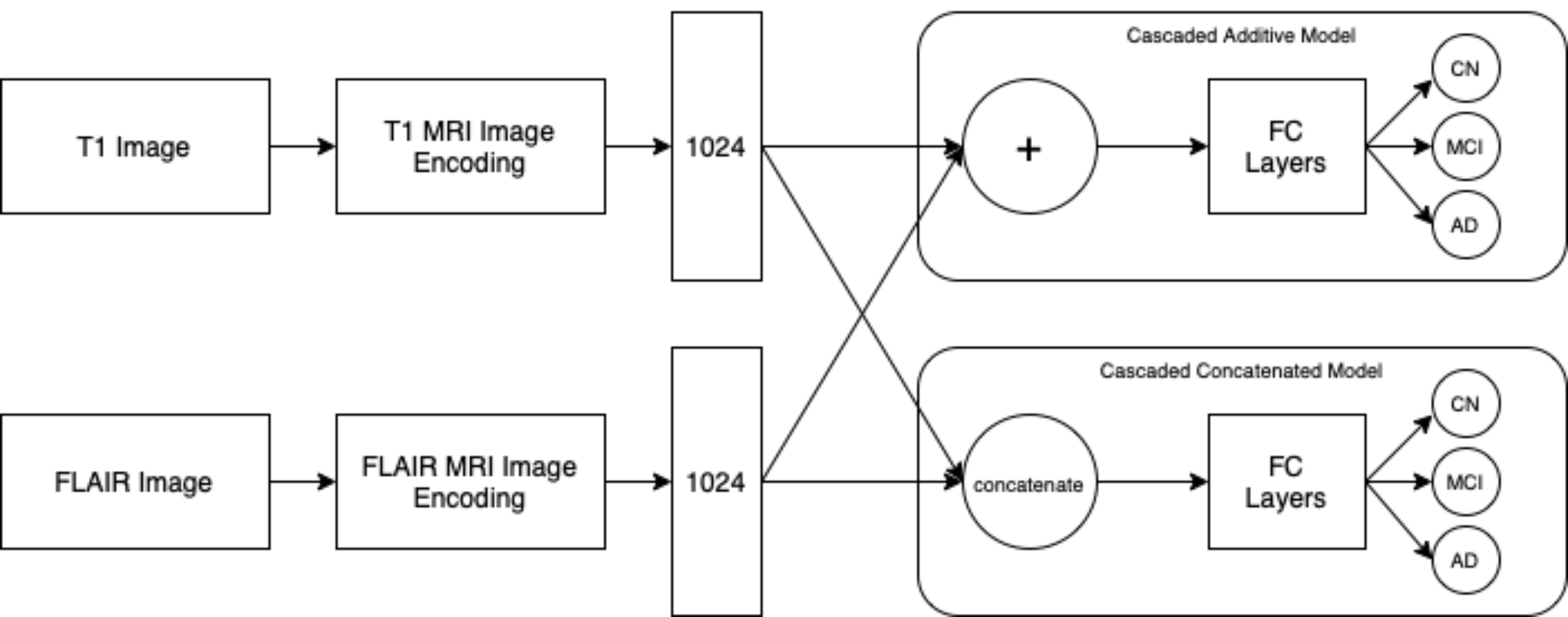}
    \caption{Model architecture of two proposed cascaded models. The dimension of input encoding for fully-connected layer is 1024 for cascaded additive model and 2048 for cascaded concatenated model. Both model utilizes a 2-layer fully-connected classifier with hidden layer size of 512.}
    \label{fig:cascaded_model_architecture}
\end{figure}

\subsubsection{T1-FLAIR transfer learning}
For the T1-FLAIR transfer learning task, we aim to check the performance of using pre-trained model with clinical T1-weighted scans onto FLAIR scans. Crossing modality, this should be a significantly harder task in comparison with T1-T1 transfer learning task. For baseline, we utilized the model trained completely with clinical T1 scans (i.e. the re-trained model from our T1-T1 transfer learning task). When then perform similar fine-tuning and re-training experiments aforementioned.

\subsubsection{Weighted multi-modal learning}
In this task, we aim to combine the prediction outputs from T1 and FLAIR models for a multi-modal prediction. We perform a simple class-wise weighting with model outputs. For each class $c$ of AD, MCI or CN with output $P_{T1}$ and $P_{FL}$ from T1 and FLAIR models, respectively, the weighted probability score is given by:
\[
\mathbb{P}_\alpha^c ~ (p_{T1}, ~ p_{FL}) = \alpha ~ p_{T1} + (1-\alpha) ~ p_{FL}.
\]

This serves as a simple, baseline multi-modal model. We would like to investigate preliminarily the feasibility of multi-modal learning under dataset setting. We find the best weight coefficient for each metric, respectively, using validation dataset.

\subsubsection{Cascaded multi-modal deep neural network}
We further investigate the feasibility of multi-modal learning with a complete deep neural network (DNN) model architecture. We propose two similar models, with the pre-trained T1 and FLAIR CNN image encodings as backbone, by adding the image encodings or by concatenating the two encodings, as shown in Figure.~\ref{fig:cascaded_model_architecture}. The only difference between additive and concatenated model is the input for the fully-connected classifier, where the additive model adds up the two image encodings (dim=1024) while the concatenated model concatenates the two image encodings into a flat emcoding (dim=2048). The fully-connected classifier has 2 layers, with hidden layer dimension set to 512. The model outputs a softmax probability for AD, MCI and CN. Using pre-trained image embeddings, we only train the final fully-connected layer, with batch size set to 4 and learning rate to 0.01. We use stochastic gradient descent with momentum equal to 0.9. We train each model for 200 epochs.

\subsection{Results}
\begin{figure}[t]
    \centering
    \begin{subfigure}{0.32\textwidth}
        \centering
        \includegraphics[width=\textwidth]{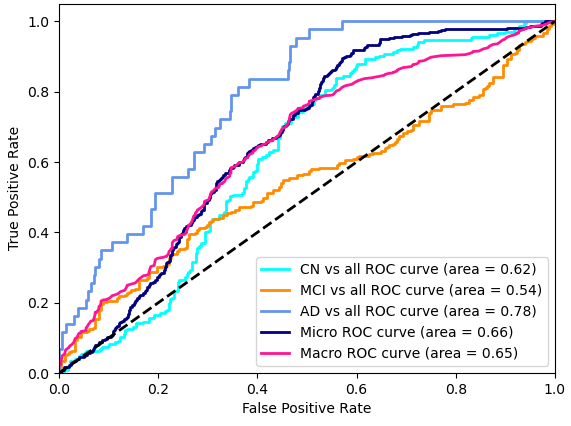}
        \par\vspace{-.2cm}
        \caption{T1 baseline}
        \label{fig:t1_baseline}
    \end{subfigure}
    \hfill
    \begin{subfigure}{0.32\textwidth}
        \centering
        \includegraphics[width=\textwidth]{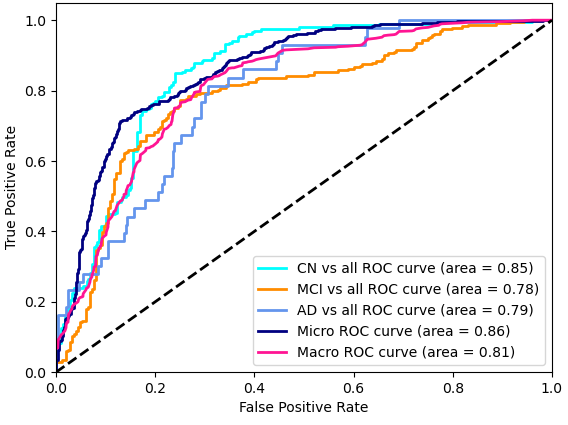}
        \par\vspace{-.2cm}
        \caption{T1 fine-tune}
        \label{fig:t1_finetune}
    \end{subfigure}
    \hfill
    \begin{subfigure}{0.32\textwidth}
        \centering
        \includegraphics[width=\textwidth]{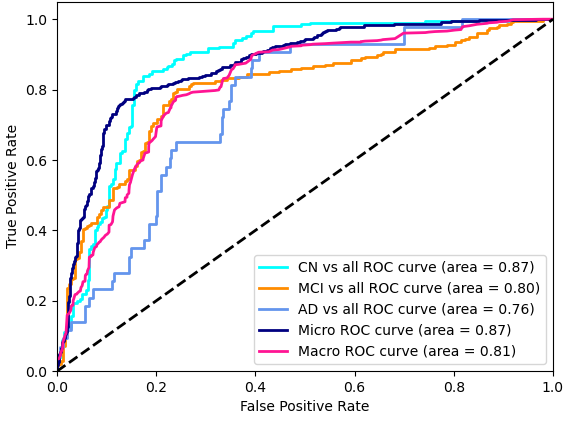}
        \par\vspace{-.2cm}
        \caption{T1 re-train}
        \label{fig:t1_retrain}
    \end{subfigure}
    \par\smallskip
    \begin{subfigure}{0.32\textwidth}
        \centering
        \includegraphics[width=\textwidth]{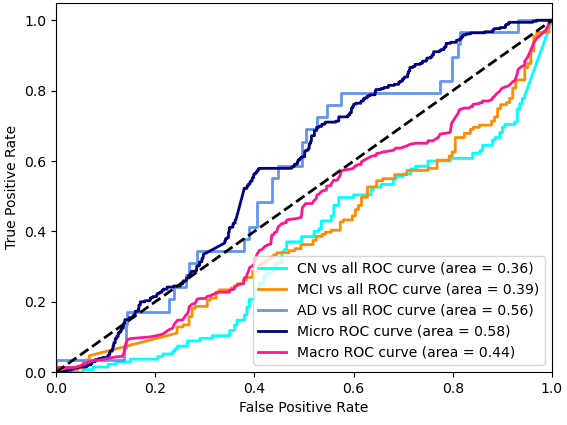}
        \par\vspace{-.2cm}
        \caption{FLAIR baseline}
        \label{fig:flair_baseline}
    \end{subfigure}
    \hfill
    \begin{subfigure}{0.32\textwidth}
        \centering
        \includegraphics[width=\textwidth]{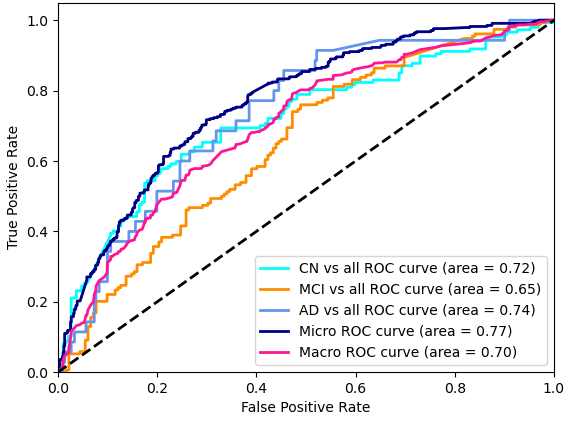}
        \par\vspace{-.2cm}
        \caption{FLAIR fine-tune}
        \label{fig:flair_finetune}
    \end{subfigure}
    \hfill
    \begin{subfigure}{0.32\textwidth}
        \centering
        \includegraphics[width=\textwidth]{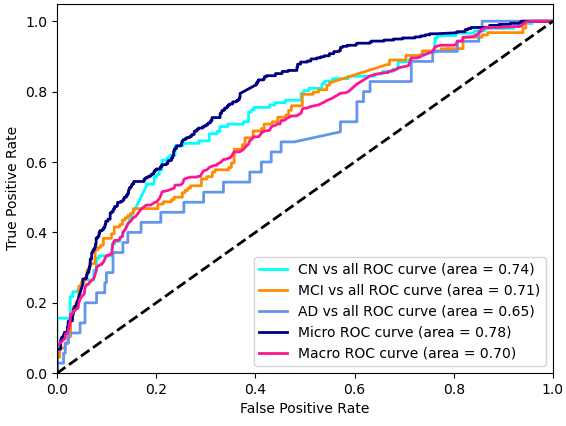}
        \par\vspace{-.2cm}
        \caption{FLAIR re-train}
        \label{fig:flair_retrain}
    \end{subfigure}
    \par\smallskip
    \begin{subfigure}{0.32\textwidth}
        \centering
        \includegraphics[width=\textwidth]{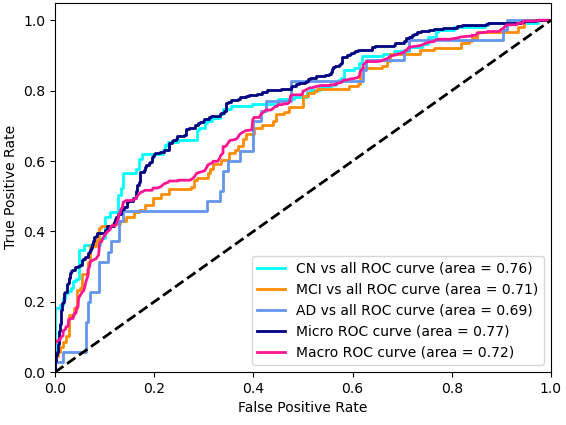}
        \par\vspace{-.2cm}
        \caption{Cascaded additive}
        \label{fig:cascaded_additive}
    \end{subfigure}
    \hspace{.3cm}
    \begin{subfigure}{0.32\textwidth}
        \centering
        \includegraphics[width=\textwidth]{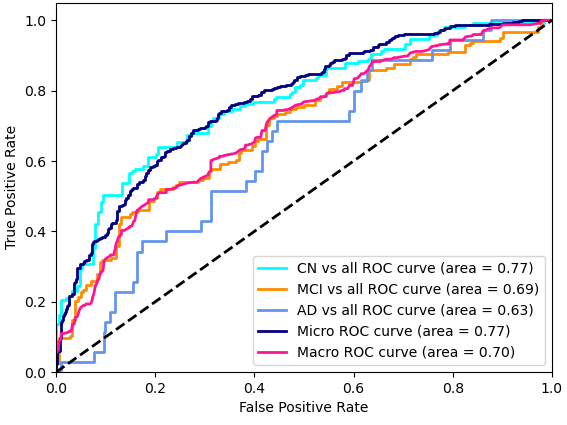}
        \par\vspace{-.2cm}
        \caption{Cascaded concatenated}
        \label{fig:cascaded_concat}
    \end{subfigure}
    \caption{ROC evaluations for transfer learning experiments. In general, baseline models perform poorly. With fine-tuning, model performance improves significantly and with re-training, model performance improves marginally.}
    \label{fig:transfer learning roc}
\end{figure}

\begin{table}[]
    \centering
    \begin{tabular}{|c||c|c|c|c|c|}
    \hline
         \textbf{Experiment}& \textbf{CN vs. All} & \textbf{MCI vs. All} & \textbf{AD vs. All} & \textbf{Micro} &  \textbf{Macro} \\
         \hline
         T1 baseline & .62 & .54 & .78 & .66 & .65 \\
         T1 fine-tune & .85 & .78 & .79 & .86 & .81 \\
         T1 re-train & .87 & .80 & .76 & .87 & .81 \\
         \hline
         FLAIR baseline & .36 & .39 & .56 & .58 & .44 \\
         FLAIR fine-tune & .72 & .65 & .74 & .77 & .70 \\
         FLAIR re-train & .74 & .71 & .65 & .78 & .70 \\
         \hline
         Multi-modal weighted* & \textbf{.89} & \textbf{.83} & \textbf{.85} & \textbf{.88} & \textbf{.86} \\
         Cascaded additive & .76 & .71 & .69 & .77 & .72 \\
         Cascaded concat & .77 & .69 & .63 & .77 & .70 \\
         \hline
    \end{tabular}
    \par\bigskip
    \caption{Model performance by AUC. Performance on T1 is significantly better than on FLAIR. Multi-modal weighted model has the best performance across all AUC evaluations. Note that AUC scores reported for multi-modal weighted model are not evaluated using one weight, instead, we utilize one optimal weight for each AUC evaluation. The cascaded DNN models are not performing well.}
    \label{tab:Model performance}
\end{table}

Results for all experiments are reported in Table.~\ref{tab:Model performance}. We first analyze T1-T1 transfer learning experiment. The baseline model yields AUC socre of CN vs All as 0.62; MCI vs All as 0.54; AD vs All as 0.78. With fine-tuning of the model, the AUC socre of CN vs All improves significantly to 0.86; MCI vs All to 0.78; AD vs All to 0.79.  Retraining the model, however, only brings marginal improvement of performance. Overall, the transfer learning based on T1 modality achieves great performance.

Next, we probe into the second task, checking whether the result model from our first task can fulfill transfer learning from T1 modality to FLAIR modality. By directly applying the model on FLAIR modality, we get the score as CN vs All equals to 0.36; MCI vs All equals to 0.39; AD vs All equals to 0.56. It's reasonable that the classifier performs poorly, considering that the vidual domain shift of T1 modality and FLAIR modality is significant. Next, we fine-tune the model, substantially improving CN vs All to 0.74; MCI vs All equals to 0.71; AD vs All equals to 0.65. This can be explained by the fact that extracted features of T1 modality can be similar to that of FLAIR modality, but requires for a better classifier to produce desired results. To sum up, the transfer learning from T1 modality to FLAIR modality looks ideal.

Evaluating on the transfer learning experiments, we first find that transfer learning experiments do not outperform trained-from-scratch models. We hypothesize the phenomenon to be related to discrepancies in T1 image qualities or attributes between datasets. Next, FLAIR models, even the retrained one, perform poorly in comparison to T1 models. We suspect this due to limitation in our FLAIR data preprocessing techniques, where the FLAIR image is up-sampled to match the template of the referencing T1 image, since FLAIR images, technically, retain lower resolution than T1 images.

\begin{figure}[t]
    \centering
    \includegraphics[width=.5\textwidth]{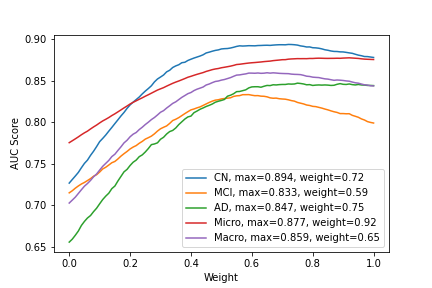}
    \caption{ROC evaluations for baseline weighted multi-modal model with respect to weight coefficients. We observe boosts in AUC-ROC score for some categories (macro AUC and the three one-vs-rest AUC sscores) under different weights. The optimal weighting shows greater importance of T1 scans over FLAIR scans.}
    \label{fig:baseline multimodal}
\end{figure}

We now examine the feasibility of multi-modal learning. To set up a baseline, we utilize trained-from-scratch T1 and FLAIR models, and investigate weighting of their predicted values as final prediction. The result of performance with different candidates of weighting coefficients are shown in Figure.~\ref{fig:baseline multimodal}. In general, model performance improves with this simple weighting technique. Observe that the optimal weights are closer to 1, indicating more importance being given to T1 scans over FLAIR scans.

Finally, the two cascaded DNN models yield disappointing performances, both with micro AUC score of .77 and macro AUC score of .72 and .70 for additive and concatenated model respectively. Such results could be explained by two issues. First, the model seems to be bounded performance-wise by the unsatisfactory FLAIR model used as embedding layer, as the two models achieve performances resemble that of the FLAIR model. The DNN models are, for some reason, unable to credit T1 modality with higher weighting, as our baseline multi-modal weighted model does. Next, although we are using encodings from the two modalities, we did not exploit the fact that the pair of T1-FLAIR input is from the same session and thus possesses inter-relationship in various aspects (e.g. position-wise).

\section{Conclusions}
Our results empirically evidence that transfer learning on dementia detection is effective and multi-modal learning does improve the detection performance. Our models have achieved competitive micro-AUC scores comparing to the state-of-the-art work on transfer learning and multi-modal learning tasks on dementia detection. We believe that our models makes accurate prediction in classifying the three stages of the dementia, and are ready to be used for deployment for real world problems. Possible settings of deployment include early detection of Alzheimer’s disease and determination of accurate treatment, and disease factor analysis. Our work is not only useful in academical research of clinical MRI and dementia detection, but also practical prevention and treatment of Alzheimer's disease. 

Due to the limitation of time, there are inevitably some shortcomings we have not addressed on. But these areas are interesting to proceed in the future. We are not yet satisfied with the current performance of the multi-modal learning models. Self-attention based models such as the multi-modal transformer proposed by \citet{dai2021transmed} are proven to work on medical images. Such architecture is a good candidate to experiment on our data set. Next, we have not performed any form of factor analysis to interpret our models. It will offer additional insights and implications for Physicians to better understand and detect dementia if we continue discovering the interpretability of the learned representation. In addition, as mentioned in previous sections, FLAIR modality are in principle different to T1 modalities in scan resolution. Future work could be done to address this problem, by innovating the image preprocessing pipeline or by designing a model architecture that better captures features in FLAIR scans.

\bibliography{ref}

\newpage

\appendix

\section{Data Preprocessing Pipelines}

\begin{figure}[h]
    \centering
    \begin{subfigure}{\textwidth}
        \centering
        \includegraphics[width=\textwidth]{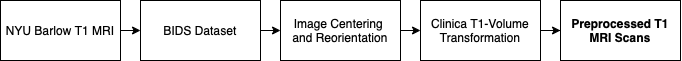}
        \caption{T1 preprocessing pipeline}
        \label{fig:t1_preprocess}
    \end{subfigure}
    \par\vspace{1cm}
    \begin{subfigure}{\textwidth}
        \centering
        \includegraphics[width=\textwidth]{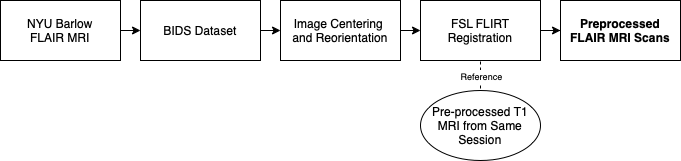}
        \caption{FLAIR preprocessing pipeline}
        \label{fig:flair_preprocess}
    \end{subfigure}
    \par\vspace{1cm}
    \begin{subfigure}{\textwidth}
        \centering
        \includegraphics[width=\textwidth]{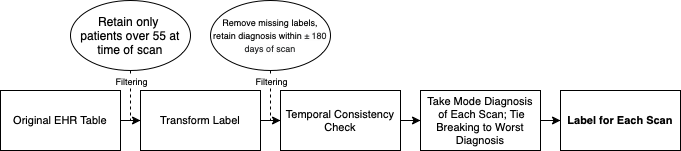}
        \caption{Labelling pipeline}
        \label{fig:label_preprocess}
    \end{subfigure}
    \caption{Image and label preprocessing pipelines.}
    \label{fig:preprocessing_pipeline}
\end{figure}

\section{Code Repository}
Our code and notebooks, excluding those with privacy concerns, are posted \href{https://github.com/jackatcl/DL_for_Dementia/tree/8866e492d517571c809fab28fddeb57931be9cc2}{here on Github}. 

\end{document}